\documentclass{article}
\usepackage{amssymb}
\usepackage{amsmath}

\usepackage[utf8]{inputenc}
\usepackage[a4paper, total={426pt, 674pt}]{geometry}
\usepackage{hyperref}

\usepackage{graphicx}
\usepackage{subcaption}
\newtheorem{theorem}{Theorem}

\usepackage{enumitem}
\date{}
\author{Philip Bille\thanks{Supported by the Danish Research Council (DFF -- 4005-00267, DFF -- 1323-00178)}\\ \texttt{phbi@dtu.dk}
    \and Anders Roy Christiansen\thanks{Supported by the Danish Research Council (DFF -- 4005-00267)}\\ \texttt{miet@dtu.dk}
    \and Mikko Berggren Ettienne$^\dag$\\ \texttt{miet@dtu.dk}
\and Inge Li G{\o}rtz$^\dag$\\ \texttt{inge@dtu.dk}}

\graphicspath{{./graphics/}}
\usepackage{tikz}

\newcommand{\aaccess}{\ensuremath\mathsf{access}}
\newcommand{\ainsert}{\ensuremath\mathsf{insert}}
\newcommand{\adelete}{\ensuremath\mathsf{delete}}
\newcommand{\aupdate}{\ensuremath\mathsf{update}}
\newcommand{\ashift}{\ensuremath\mathsf{shift}}

\newcommand{\offset}{\ensuremath\mathrm{off}}
\newcommand{\capacity}{\ensuremath\mathrm{cap}}

\newcommand{\height}{\ensuremath\mathrm{height}}
\newcommand{\elements}{\ensuremath\mathrm{elems}}

\title{Fast Dynamic Arrays}

\begin{document}

\maketitle

\abstract{We present a highly optimized implementation
    of tiered vectors, a data structure
    for maintaining a sequence of $n$ elements
    supporting access in time $O(1)$
    and insertion and deletion in time $O(n^\epsilon)$ for $\epsilon > 0$
    while using $o(n)$ extra space.
    We consider several different implementation optimizations 
    in C++ and compare their performance to that of \text{vector} and
    \text{multiset}
    from the standard library on sequences with up to
    $10^8$ elements.
    Our fastest implementation uses
    much less space than multiset 
    while providing speedups of $40\times$ for
    access operations compared to multiset
    and speedups of $10.000\times$ compared
    to vector for insertion and deletion operations
    while being competitive
    with both data structures for all other
    operations.
}

\section{Introduction}

We present a highly optimized implementation
of a data structure solving the \textit{dynamic array problem},
that is, maintain a sequence of elements subject to the following operations:

\begin{description}
    \item[$\quad\aaccess(i)$:] return the $i^{th}$ element
        in the sequence.
	\item[$\quad\aaccess(i, m)$:] return the $i^{th}$ through $(i+m-1)^{th}$
            elements in the sequence.
	\item[$\quad\ainsert(i, x)$:] insert element $x$ 
            immediately after the $i^{th}$ element.
	\item[$\quad\adelete(i)$:] remove the
            $i^{th}$ element from the sequence.
	\item[$\quad\aupdate(i, x)$:] exchange the $i^{th}$ element
            with $x$.
\end{description}
\noindent
This is a fundamental and well studied
data structure problem~\cite{Dietz:1989:OAL:645928.672529,
    Fredman:1989:CPC:73007.73040, Katajainen2001,
    Raman2001, Frederickson:1983:IDS:322358.322364,
    Brodnik:1999:RAO:645932.673194,
    Goodrich1999,Katajainen2016}
solved by textbook data structures like arrays and binary trees.
Many dynamic trees provide all the operations in
$O(\lg n)$ time including 2-3-4 trees, AVL trees, splay trees, etc.\
and Dietz~\cite{Dietz:1989:OAL:645928.672529} gives
a data structure that matches the lower bound
of $\Omega(\lg n/ \lg \lg n)$  showed by
Fredman and Saks~\cite{Fredman:1989:CPC:73007.73040}.
The lower bound only holds when identical complexities are required for all operations.
In this paper we focus on the variation where $\aaccess$ must run in $O(1)$ time.
Goodrich and Kloss present what they call \textit{tiered vectors}
\cite{Goodrich1999} with a time complexity of $O(1)$ 
for $\aaccess$ and $\aupdate$ and $O(n^{1/l})$ for $\ainsert$ and $\adelete$
for any constant integer $l \geq 2$, using ideas similar 
to Frederickson's in~\cite{Frederickson:1983:IDS:322358.322364}.
The data structure
uses only $o(n)$ extra space beyond that required to store the actual elements. 
At the core, the data structure is a tree with out degree
$n^{1/l}$ and \emph{constant} height $l - 1$.

Goodrich and Kloss compare the performance
of an implementation with $l = 2$ to that of
\textit{vector} from the standard library of Java
and show that the structure is competitive for
access operations while being significantly faster
for insertions and deletions.
Tiered vectors provide 
a performance trade-off between standard arrays and balanced binary trees
for the dynamic array problem.

\subparagraph{Our Contribution}

In this paper, we present what we believe is the first implementation of tiered vectors that supports more than 2 tiers. Our C++ implementation supports $\aaccess$ and $\aupdate$ in times that are competitive with the vector data structure from C++'s standard library while $\ainsert$ and $\adelete$ run more than $10.000\times$ faster. It performs $\aaccess$ and $\aupdate$ more than $40 \times$ faster than the multiset data structure from the standard library while $\ainsert$ and $\adelete$ is only a few percent slower. Furthermore multiset uses more than $10\times$ more space than our implementation. All of this when working on large sequences of $10^8$ 32-bit integers.

To obtain these results, we significantly decrease the number of memory probes per operation compared to the original tiered vector. Our best variant requires only half as many memory probes as the original tiered vector for $\aaccess$ and $\aupdate$ operations which is critical for the practical performance. Our implementation is cache efficient which makes all operations run fast in practice even on tiered vectors with several tiers.

We experimentally compare the different variants of tiered vectors. Besides the comparison to the two commonly used C++ data structures, vector and multiset, we compare the different variants of tiered vectors to find the best one. We show that the number of tiers have a significant impact on the performance which underlines the importance of tiered vectors supporting more than 2 tiers.

Our implementations are parameterized and thus support any number of tiers
$\geq 2$. We use techniques like \textit{template recursion} to keep the code rather
simple while enabling the compiler to generate highly optimized code.

The source code can be found at \url{https://github.com/mettienne/tiered-vector}.

\section{Preliminaries}

The first and $i^{th}$ element of a sequence $A$ are denoted 
$A[0]$ and $A[i-1]$ respectively
and the $i^{th}$ through $j^{th}$
elements are denoted $A[i-1,j-1]$. Let $A_1 \cdot A_2$ denote the concatenation
of the sequences $A_1$ and $A_2$. $|A|$ denotes the number of elements in the
sequence $A$. A circular shift of a sequence $A$ by $x$ is the sequence
 $A[|A| - x, |A| - 1] \cdot A[0, |A| - x - 1]$. Define the remainder of division of $a$
by $b$ as $a \mod b = a - qb$ where $q$ is the largest integer such that $q \cdot b \leq a$.
Define $A[i,j] \mod w$ to be the elements $A[i \mod w], A[(i+1) \mod w], \ldots , A[j\mod w]$,
i.e. $A[4,7] \mod 5$ = $A[4],A[0],A[1],A[2]$. Let $\lfloor x \rfloor$ denote the largest integer smaller than $x$.

\section{Tiered Vectors}

In this section we will describe how the tiered vector data structure
from~\cite{Goodrich1999} works. 

\begin{figure}
	\includegraphics[width=\textwidth]{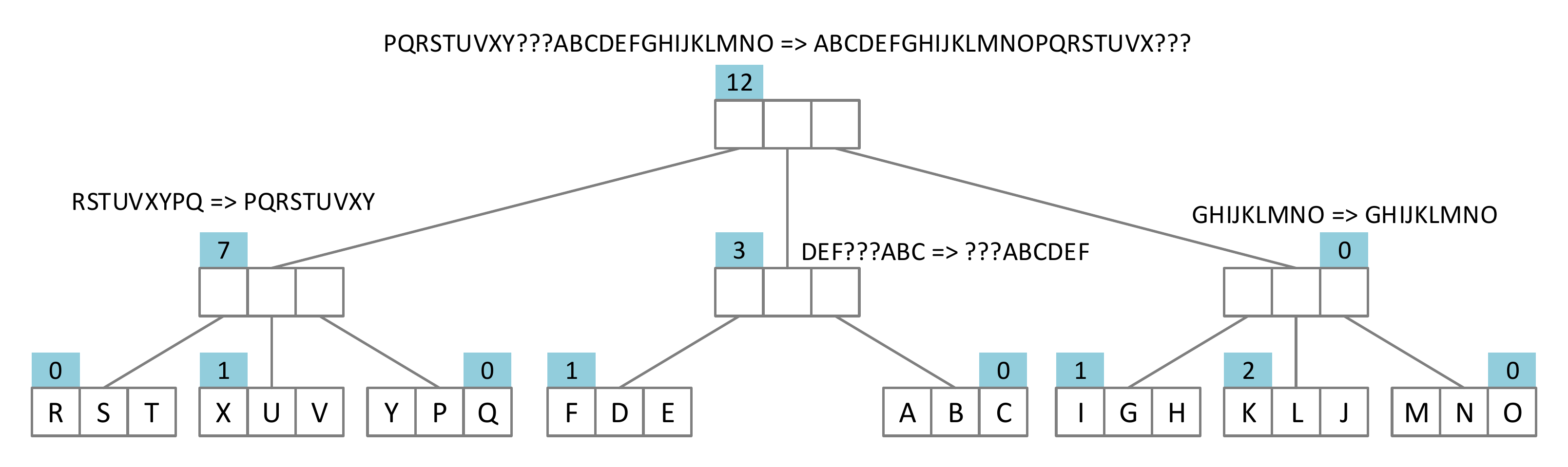}
    \caption{An illustration of a tiered vector with $l = w = 3$. The elements are letters, and the tiered vector represents the sequence ABCDEFGHIJKLMNOPQRSTUVX. The elements in the leaves are the elements that are actually stored. The number above each node is its offset. The strings above an internal node $v$ with children $c_1, c_2, c_3$ is respectively $A(c_1) \cdot A(c_2) \cdot A(c_3)$ and $A(v)$, i.e.\ the elements $v$ represents before and after the circular shift. ? specifies an empty element.}
\label{fig:ds}
\end{figure}

%\paragraph{Data Structure} 
\subparagraph*{Data Structure} 
An $l$-tiered vector can be seen as a tree $T$ with root $r$, fixed
height $l - 1$ and out-degree $w$ for any $l \geq 2$.
A node $v \in T$ represents a sequence of elements $A(v)$ thus 
$A(r)$ is the sequence represented by the tiered vector. The capacity $\capacity(v)$ of a node $v$ is $w^{\height(v)+1}$. For a node $v$ with children $c_1, c_2, \ldots, c_w$, $A(v)$ is a circular shift of the
concatenation of the elements represented by its children, 
$A(c_1) \cdot A(c_2) \cdot \ldots \cdot A(c_w)$.
The circular shift is determined by an integer $\offset(v)
\in [\capacity(v)]$ that is explicitly stored for all nodes. Thus the sequence of
elements $A(v)$ of an internal node $v$ can be reconstructed by recursively
reconstructing the sequence for each of its children, concatenating these and
then circular shifting the sequence by $\offset(v)$. See Figure~\ref{fig:ds} for an illustration. A leaf $v$ of $T$
explicitly stores the sequence $A(v)$ in a circular array $\elements(v)$ with
size $w$ whereas internal nodes only store their respective offset.
 Call a node $v$ full if $|A(v)| = \capacity(v)$ and empty if $|A(v)| = 0$. In order to support fast $\aaccess$, for all nodes $v$ the elements of $A(v)$ are located in consecutive children of $v$ that
are all full, except the children containing the first
and last element of $A(v)$ which may be only partly full.

%\paragraph{Access \& Update}
\subparagraph*{Access \& Update}
To access an element $A(r)[i]$ at a given index $i$; one traverses a
path from the root down to a leaf in the tree. In each node the offset of the
node is added to the index to compensate for the cyclic shift, and the traversing is continued in the child corresponding to the newly calculated index. 
Finally when reaching a leaf, the desired element is
returned from the elements array of that leaf. The operation $\aaccess(v, i)$ returns the
element $A(v)[i]$ and is recursively computed as follows:

\begin{description} 
    \item[\quad v is internal:] Compute $i' = (i + \offset(v))
        \mod \capacity(v)$, let $v'$ be the $\lfloor i' / w \rfloor^{th}$ child of $v$ and return the element
    $\aaccess(v', i' \mod \capacity(v'))$. 
    
\item[\quad v is leaf:] Compute $i' = (i + \offset(v)) \mod w$
    and return the element $\elements(v)[i']$.
    \end{description}

The time complexity is $\Theta(l)$ as we visit all nodes on a root-to-leaf path in $T$. To navigate this path we must follow $l - 1$ child pointers, lookup $l$ offsets, and access the element itself. Therefore this requires $l - 1 + l + 1 = 2l$ memory probes.

The update operation is entirely similar to access, except
the element found is not returned but substituted with the new element. The
running time is therefore $\Theta(l)$ as well. For future use, let $\aupdate(v, i, e)$ be the operation that sets $A(v)[i] = e$ and returns the element that was
substituted. 

%\paragraph{Range Access}
\subparagraph*{Range Access}

Accessing a range of elements, can obviously be done by using the
$\aaccess$-operation multiple times, but this results in redundant traversing
of the tree, since consecutive elements of a leaf often
-- but not always due to circular shifts -- corresponds to consecutive elements of $A(r)$.
Let $\aaccess(v, i, m)$ report the
elements $A(v)[i \ldots i + m - 1]$ in order. The operation can recursively
be defined as:

\begin{description} \item[\quad v is internal:] 
    Let $i_l = (i + \offset(v)) \mod \capacity(v)$,
    and let $i_r = (i_l + m) \mod \capacity(v)$. The children of
    $v$ that contains the elements to be reported are in the range $[\lfloor i_l \cdot w / \capacity(v) \rfloor, \lfloor i_r \cdot w / \capacity(v) \rfloor] \mod w$,
    call these $c_l, c_{l+1},
    \ldots, c_r$. In order, call $\aaccess(c_l, i_l, \min(m, \capacity(c_l) -
    i_l))$, $\aaccess(c_i, 0, \capacity(c_i))$ for $c_i = c_{l+1}, \ldots,
    c_{r-1}$, and $\aaccess(c_r, e_{r-1}, 0, i_r \mod \capacity(c_r))$.
	
        \item[\quad v is leaf:] Report the elements $\elements(v)[i, i+m-1] \mod w$. \end{description}

The running time of this strategy is $O(lm)$, but saves a constant factor over the naive solution.

%\paragraph{Insert \& Delete}
\subparagraph*{Insert \& Delete}

Inserting an element in the end (or beginning) of the array can simply be
achieved using the $\aupdate$-operation. Thus the interesting part is fast insertion at an arbitrary position;
this is where we utilize the offsets.

Consider a node $v$, the key challenge is to shift a big chunk of elements $A(v)[i, i+m-1]$ one index right (or left) to $A(v)[i+1, i+m]$ to make room for a new element (without actually moving each element in the range). Look at the range of children $c_l, c_{l+1}, \ldots, c_r$ that covers the range of elements $A(v)[i, i+m-1]$ to be shifted. All elements in $c_{l+1}, \ldots, c_{r-1}$ must be shifted. These children are guaranteed to be full, so make a circular shift by decrementing each of their offsets by one. Afterwards take the element $A(c_{i-1})[0]$ and move it to $A(c_{i})[0]$ using the $\aupdate$ operation for $l < i \leq r$. In $c_l$ and $c_r$ only a subrange of the elements might need shifting, which we do recursively. In the base case of this recursion, namely when
$v$ is a leaf, shift the elements by actually moving the elements one-by-one in $\elements(v)$.

Formally we define the $\ashift(v, e, i, m)$ operation that (logically) shifts
all elements $A(v)[i, i+m-1]$ one place right to $A[i+1, i+m]$, sets $A[i] = e$ and returns the value that was previously on position $A[i+m]$ as:

\begin{description} \item[\quad v is internal:] Let $i_l = (i + \offset(v)) \mod
    \capacity(v)$, and let $i_r = (i_l + m) \mod \capacity(v)$. The children of
    $v$ that must be updated are in the range $[\lfloor i_l \cdot w / \capacity(v) \rfloor, \lfloor i_r \cdot w / \capacity(v) \rfloor] \mod w$ call these $c_l, c_{l+1}, \ldots, c_r$.
Let $e_l = \ashift(c_l, e, i_l, \min(m, \capacity(c_l) - i_l))$. Let $e_i =
\aupdate(c_i, size(c) - 1, e_{i-1})$ and set $\offset(c_i) = (\offset(c_i) - 1)
\mod \capacity(c)$ for $c_i = c_{l+1}, \ldots, c_{r-1}$. Finally call
$\ashift(c_r, e_{r-1}, 0, i_r \mod \capacity(c_r))$.
	
        \item[\quad v is leaf:] Let $e_o = \elements(v)[(i+m) \mod w]$. Move the
            elements $\elements(v)[i, (i+m-1) \mod w]$ to $\elements(v)[i+1,(i+m) \mod w]$, and set $\elements(v)[i] = e$. Return $e_o$.
    \end{description}

An insertion $\ainsert(i, e)$ can then be performed as $\ashift(root, e, i,
size(root) - i - 1)$. The running time of an insertion is $T(l) = 2T(l - 1) + w\cdot l \Rightarrow T(l) = O(2^l w)$.

%TODO: {Add illustration.}

A deletion of an element can basically be done as an inverted insertion, thus
deletion can be implemented using the $\ashift$-operation from before. A
$\adelete(i)$ can be performed as $\ashift(r, \bot, 0, i)$ followed by an
update of the root's offset to $(\offset(r) + 1) \mod \capacity(r)$.

%\paragraph{Space}
\subparagraph*{Space}

There are at most $O(w^{l-1})$ nodes in the tree and each takes up constant
space, thus the total space of the tree is $O(w^{l-1})$.
All leaves are either empty or full except the two leaves storing the first and
last element of the sequence which might contain less than $w$ elements.
Because the arrays of empty leaves are not allocated the space overhead of the arrays is $O(w)$.
Thus beyond the space required to store the $n$ elements themselves, tiered vectors
have a space overhead of $O(w^{l-1})$.

To obtain the desired bounds $w$ is maintained such that $w = \Theta(n^\epsilon)$ where $\epsilon = 1/l$ and $n$ is the number of elements in the tiered vector. This can be achieved by using global rebuilding to gradually increase/decrease the value of $w$ when elements are inserted/deleted without asymptotically changing the running times. We will not provide the details here. We sum up the original tiered vector data structure in the following theorem:

\begin{theorem}[\cite{Goodrich1999}] The original $l$-tiered vector solves the
    dynamic array problem for $l \geq 2$ using $\Theta(n^{1-1/l})$ extra space
    while supporting $\aaccess$ and $\aupdate$ in $\Theta(l)$ time and $2l$
    memory probes. The operations $\ainsert$ and $\adelete$ take $O(2^l n^{1/l})$ time.
    \label{thm:pointer}
\end{theorem}

\section{Improved Tiered Vectors}

In this paper, we consider several new variants of the tiered vector. This section considers the theoretical
properties of these approaches. In particular we are interested in the number of memory accesses that are required for the different memory layouts, since this turns out to have an effect on the experimental running time.
In Section~\ref{sec:experimental} we analyze the actual impact in practice through experiments.

\subsection{Implicit Tiered Vectors}

As the degree of all nodes is always fixed at some constant value $w$ (it may be
changed for all nodes when the tree is rebuilt due to a full root), it is possible to layout the
offsets and elements such that no pointers are necessary to navigate the
tree. Simply number all nodes from left-to-right level-by-level starting in the
root with number 0. Using this numbering scheme, we can store all
offsets of the nodes in a single array and similarly all the elements of the leaves in another array.

To access an element, we only have to lookup the offset for each
node on the root-to-leaf path which requires $l-1$ memory probes plus the final
element lookup, i.e.\ in total $l$ which is half as many as
the original tiered vector.
 The downside with this representation is that it must allocate the
two arrays in their entirety at the point of initialization (or when rebuilding). This results in a $\Theta(n)$ space overhead which is worse than the $\Theta(n^{1-\epsilon})$ space overhead from the original tiered vector.

\begin{theorem} The implicit $l$-tiered vector solves the dynamic array problem for $l \geq 2$
using $O(n)$ extra space while supporting $\aaccess$ and $\aupdate$ in
$O(l)$ time requiring $l$ memory probes. The operations $\ainsert$ and
$\adelete$ take $O(2^l n^{1/l})$ time.
\label{thm:implicit}
\end{theorem}

\subsection{Lazy Tiered Vectors}

We now combine the original and the implicit representation, to get both few memory probes and little space overhead. Instead of having a single array storing all
the elements of the leaves, we store for each leaf a pointer to a location with an array containing the leaf's elements. The array is lazily allocated in memory when elements are actually inserted into it.

The total size of the offset-array and the element pointers in the leaves is $O(n^{1-\epsilon})$. At most two leaves are only partially full, therefore the
total space is now again reduced to $O(n^{1-\epsilon})$. To navigate a root-to-leaf path, we now need to look at $l - 1$ offsets, follow a pointer from a leaf to its array and access the element in the array, giving a total of $l + 1$
memory accesses.

\begin{theorem}
        The lazy $l$-tiered vector solves the dynamic array problem for $l \geq 2$ using
        $\Theta(n^{1-1/l})$ extra space while supporting $\aaccess$ and
        $\aupdate$ in $\Theta(l)$ time requiring $l+1$ memory probes. The
        operations $\ainsert$ and $\adelete$ take $O(2^l n^{1/l})$
        time.
\label{thm:lazy}
\end{theorem}

\section{Implementation}

\label{sec:implementation}
We have implemented a generic version of the tiered vector data structure such
that the number of tiers and the size of each tier can be specified at compile
time. To the best of our knowledge, all prior implementations of the tiered vector are limited to the considerably simpler 2-tier version.
Also, most of the performance optimizations applied in the 2-tier
implementations do not easily generalize.
We have implemented the following variants of tiered vectors:
%\addtolength\leftmargini{7.5em}
%\begin{enumerate}[itemsep=-1ex,partopsep=1ex,parsep=1ex, itemindent=1.5em,align=left]
\begin{itemize}

    %\item[\textit{Original}]
    \item \textit{Original}
        The data structure described in Theorem~\ref{thm:pointer}.

    %\item[\textit{Optimized Original}]
    \item \textit{Optimized Original}
        As described in Theorem~\ref{thm:pointer}
        but with the offset of a node $v$ located
        in the parent of $v$, adjacent in memory to the
        pointer to $v$. Leaves only consist of an array of elements
        (since their parent store their offset)
        and the root's offset is maintained separately
        as there is no parent to store it in.

    %\item[\textit{Implicit}]
    \item \textit{Implicit}
        This is the data structure described in Theorem~\ref{thm:implicit}
        where the tree is represented implicitly in an array
        storing the offsets and the elements of the leaves are
        located in a single array.

    %\item[\textit{Packed Implicit}]
    \item\textit{Packed Implicit}
        This is the data structure described in Theorem~\ref{thm:implicit}
        with the following optimization;
        The offsets stored in the offset array
        are packed together and stored in as little space as possible.
        The maximum offset of a node $v$ in the tree is
        $n^{\epsilon(\height(v)+1)}$ and the
        number of bits needed to store all the offsets is therefore $\sum_{i=0}^l
        n^{1-i\epsilon}  \log(n^{i\epsilon}) = \log(n) \sum_{i=0}^l i\epsilon
        n^{1-i\epsilon}  \approx \epsilon n^{1-\epsilon} \log(n)$ (for sufficiently large $n$).
        Thus the $n^{1-\epsilon}$ offsets can
        be stored in approximately $\epsilon n^{1-\epsilon}$ words giving a space reduction
        of a constant factor $\epsilon$.
        The smaller memory footprint could lead to better cache performance.

    %\item[\textit{Lazy}]
    \item \textit{Lazy}
        This is the data structure described in Theorem~\ref{thm:lazy}
        where the tree is represented implicitly in an array
        storing the offsets and every leaf stores a pointer to an array
        storing only the elements of that leaf.

    %\item[\textit{Packed Lazy}]
    \item \textit{Packed Lazy}
        This is the data structure described in Theorem~\ref{thm:lazy}
        with the following optimization;
        The offset and the pointer stored in a leaf
        is packed together and stored at the same memory location.
        On most modern 64-bit systems -- including the one we are testing on
        -- a memory pointer is only allowed to address 48 bits.
        This means we have room to pack a 16 bit offset
        in the same memory location as the elements pointer,
        which results in one less memory probe during an access operation.

    %\item[\textit{Non-Templated}]
    \item \textit{Non-Templated}
            The implementations described above all use C++ templating
            for recursive functions
            in order to let the compiler do significant code optimizations.
            This implementation is template free and serves as a baseline
            to compare the performance gains given by templating.

    \end{itemize}
In Section~\ref{sec:experiments} we compare the performance of
these implementations.

\subsection{C++ Templates}

We use templates to support storing different types of data in our tiered
vector similar to what most other general purpose data structures in C++ do.
This is a well-known technique which we will not describe in detail.

However, we have also used \textit{template recursion} which is basically like a normal recursion except that the recursion parameter must be a compile-time constant. This allows the compiler to unfold the recursion at compile-time eliminating all (recursive) function calls by inlining code, and allows better local code optimizations. In our case, we exploit that the height of a tiered vector is constant.

To show the rather simple code resulting from this approach (disregarding the template stuff itself), we have included a snippet of the internals of our access operation:

\begin{verbatim}

template <class T, class Layer>
struct helper {
        static T& get(size_t node, size_t idx) {
            idx = (idx + get_offset(node)) % Layer::capacity;
            auto child = get_child(node, idx / Layer::child::capacity);
            return helper<T, typename Layer::child>::get(child, idx);
        }
}

template <class T, size_t W>
struct helper<T, Layer<W, LayerEnd> > {
        static T& get(size_t node, size_t idx) {
            idx = (idx + get_offset(node)) % L::capacity;
            return get_elem(node, idx);
        }
}
\end{verbatim}

We also briefly show how to use the data structure. To specify the desired height of the tree, and the width of the nodes on each tier, we also use templating:

\begin{verbatim}
Tiered<int, Layer<8, Layer<16, Layer<32>>>> tiered;
\end{verbatim}

This will define a tiered vector containing integers with three tiers. The
height of the underlying tree is therefore 3 where the root has 8 children,
each of which has 16 children each of which contains 32 elements. We call this configuration 8-16-32.

In this implementation of tiered vectors we have decided to let
the number of children on each level be a fixed number as described above. This
imposes a maximum on the number of elements that can be inserted. However, in a
production ready implementation, it would be simple to make it grow-able by
maintaining a single growth factor that should be multiplied on the number
of children on each level. This can be combined with the templated solution
since the growing is only on the number of children and not the height of the
tree (per definition of tiered vectors the height is constant). This will
obviously increase the running time for operations when growing/shrinking is
required, but will only have minimal impact on all other operations (they will
be slightly slower because computations now must take the growth factor into
account).

In practice one could also, for many uses, simply pick the number of children
on each level sufficiently large to ensure the number of elements that will be
inserted is less than the maximum capacity. This would result in a memory
overhead when the tiered vector is almost empty, but by choosing the right
variant of tiered vectors and the right parameters this overhead would in many
cases be insignificant.

\label{sec:experimental}

\section{Experiments}

\label{sec:experiments}
In this section we compare the tiered vector to some widely used C++ standard library containers. 
We also compare different variants of the tiered vector. 
We consider how the different representations of the data
structure listed in Section~\ref{sec:implementation}, 
and also how the height of tree and the capacity of the leaves affects the running time.
The following describes the test setup:

\subparagraph{Environment}

All experiments have been performed on a Intel Core i7-4770 CPU @ 3.40GHz with
32 GB RAM. The code has been compiled with GNU GCC version 5.4.0 with flags
``-O3''. The reported times are an average over 10 test runs.
 
 \subparagraph{Procedure}
%have been added to the data structure in
 
In all tests $10^8$ 32-bit integers 
are inserted in the data structure as a preliminary step
to simulate that it has already been
used\footnote{In order to minimize the overall running time of the experiments,
the elements were not added randomly, but we show this does not give our data
structure any benefits}.
For all the access and successor operations $10^9$ elements have been accessed
and the time reported is the average time per element.
For range access, 10.000 consecutive elements are accessed.
For insertion/deletion $10^6$ elements
have been (semi-)randomly\footnote{In order to not impact timing, a simple
access pattern has been used instead of a normal pseudo-random generator.}
added/deleted, though in the case of ``vector'' only 10.000 elements were
inserted/deleted to make the experiments terminate in reasonable time. 

\subsection{Comparison to C++ STL Data Structures}

In the following we have compared our best performing tiered vector (see the next sections) to the vector and
the multiset class from the C++ standard library.
The vector data structure directly supports the
operations of a dynamic array. The multiset class is implemented as a red-black
tree and is therefore interesting to compare with our data structure.
Unfortunately, multiset does not directly support the operations of a dynamic
array (in particular it has no notion of positions of elements). To simulate an
access operation we instead find the successor of an element in the multiset.
This requires a root-to-leaf traversal of the red-black tree, just as an access
operation in a dynamic array implemented as a red-black tree would. Insertion
is simulated as an insertion into the multiset, which again requires the same
computations as a dynamic array implemented as a red-black tree would.

Besides the random access, range access and insertion,
we have also tested the operations \textit{data dependent access},
insertion in the end, deletion, and \textit{successor} queries. In the
\textit{data dependent access} tests, the next index to lookup depends on the values of the prior
lookups. This ensures that the CPU cannot successfully pipeline
consecutive lookups, but must perform them in sequence. We test insertion in the end, since
this is a very common use case. Deletion is performed by deleting elements at
random positions. The $successor$ queries returns the successor of an element
and is not actually part of the
dynamic array problem, but is included since it is a commonly used operation on
a multiset in C++. It is simply implemented as a binary search over the elements in
both the vector and tiered vector tests where the elements are now inserted in sorted order. 

The results are summarized in Table~\ref{tab:test_comp} which shows that the vector performs slightly better than the tiered vector on all access and successor tests. As expected from the $\Theta(n)$ running time, it performs extremely poor on random insertion and deletion. For insertion in the end of the sequence, vector is also slightly faster than the tiered vector. The interesting part is that even though the tiered vector requires several extra memory lookups and computations, we have managed to get the running time down to less than the double of the vector for access, even less for data dependent access and only a few percent slowdown for range access. As discussed earlier,
this is most likely because the entire tree structure (without the elements)
fits within the CPU cache, and because the computations required has been minimized.

Comparing our tiered vector to multiset, we would expect access operations to be
faster since they run in $O(1)$ time compared to $O(\log n)$. On the other
hand, we would expect insertion/deletion to be significantly slower since it
runs in $O(n^{1/l})$ time compared to $O(\log n)$ (where $l = 4$ in these tests). We
see our expectations hold for the access operations where the tiered vector is faster by more than an order of magnitude.
In random insertions however,  the tiered vector is only $8\%$ slower -- even when operating on 100.000.000 elements. Both the tiered
vector and set requires $O(\log n)$ time for the successor operation. In our
experiments the tiered vector is 3 times faster for the successor operation.

Finally, we see that the memory usage of vector and tiered vector is almost identical.
This is expected since in both cases the space usage is dominated by the space taken by the actual elements.
The multiset uses more than 10 times as much space, so this is also a considerable drawback of the red-black tree behind this structure. 

To sum up, the tiered vectors performs better than multiset on all tests
but insertion, where it performs only slightly worse.

%\caption{Figures (a) through (e) show the performance of \textit{Tiered Arrays} (\protect\purple) compared
%to the \textit{set} (\protect\green) and \textit{vector} (\protect\blue) data structures from the C++ standard library.} \label{fig:animals}
\begin{table}
	\centering
	\begin{tabular}{|l|r|r|r|r|r|}
		\hline
		& \multicolumn{1}{l|}{\textit{tiered vector}} & \multicolumn{1}{l|}{\textit{set}} & \multicolumn{1}{l|}{\textit{set / tiered}} & \multicolumn{1}{l|}{\textit{vector}} & \multicolumn{1}{l|}{\textit{vector / tiered}} \\ \hline
		access     & $34.07$ ns                                  & $1432.05$ ns                      & 42.03                                      & $21.63$ ns                           & 0.63                                          \\ \hline
		dd-access    & $99.09$ ns                                  & $1436.67$ ns                      & 14.50                                      & $79.37$ ns                           & 0.80                                          \\ \hline
		range access   & $0.24$ ns                                   & $13.02$ ns                        & 53.53                                      & $0.23$ ns                            & 0.93                                          \\ \hline
		insert   & $1.79$ $\mu$s                               & $1.65$ $\mu$s                     & 0.92                                       & $21675.49$ $\mu$s                     & 12082.33                                      \\ \hline
		insertion in end     & $7.28$ ns                               & $242.90$ ns                     & 33.38                                       & $2.93$ ns                     & 0.40                                      \\ \hline
		successor & $0.55$ $\mu$s                               & $1.53$ $\mu$s                     & 2.75                                       & $0.36$ $\mu$s                        & 0.65                                          \\ \hline
		delete     & $1.92$ $\mu$s                               & $1.78$ $\mu$s                     & 0.93                                       & $21295.25$ $\mu$s                     & 11070.04                                      \\ \hline
		memory     & $408$ MB                               & $4802$ MB                     & 11.77                                       & $405$ MB                    & 0.99                                      \\ \hline
	\end{tabular}
	\caption{The table summarizes the performance of the implicit tiered vector
		compared to the performance of multiset and vector from the C++ standard library.\
		dd-access refers to data dependent access.}
\label{tab:test_comp}
\end{table}

\definecolor{cpurple}{RGB}{131,24,197}
\definecolor{cgreen}{RGB}{70,156,118}
\definecolor{cblue}{RGB}{11,178,228}
\definecolor{cdblue}{RGB}{11,112,173}
\definecolor{corange}{RGB}{219,162,55}
\definecolor{cyellow}{RGB}{238,228,98}
\definecolor{cred}{RGB}{110,55,38}
\newcommand{\purple}{\raisebox{2pt}{\tikz{\draw[cpurple,solid,line width=1.9pt](0,0) -- (3mm,0);}}}
\newcommand{\green}{\raisebox{2pt}{\tikz{\draw[cgreen,solid,line width=1.9pt](0,0) -- (3mm,0);}}}
\newcommand{\blue}{\raisebox{2pt}{\tikz{\draw[cblue,solid,line width=1.9pt](0,0) -- (3mm,0);}}}
\newcommand{\dblue}{\raisebox{2pt}{\tikz{\draw[cdblue,solid,line width=1.9pt](0,0) -- (3mm,0);}}}
\newcommand{\orange}{\raisebox{2pt}{\tikz{\draw[corange,solid,line width=1.9pt](0,0) -- (3mm,0);}}}
\newcommand{\yellow}{\raisebox{2pt}{\tikz{\draw[cyellow,solid,line width=1.9pt](0,0) -- (3mm,0);}}}
\newcommand{\red}{\raisebox{2pt}{\tikz{\draw[cred,solid,line width=1.9pt](0,0) -- (3mm,0);}}}

\begin{figure}[ht]
	\centering
	\begin{subfigure}[b]{0.3\textwidth}
		\includegraphics[width=\textwidth]{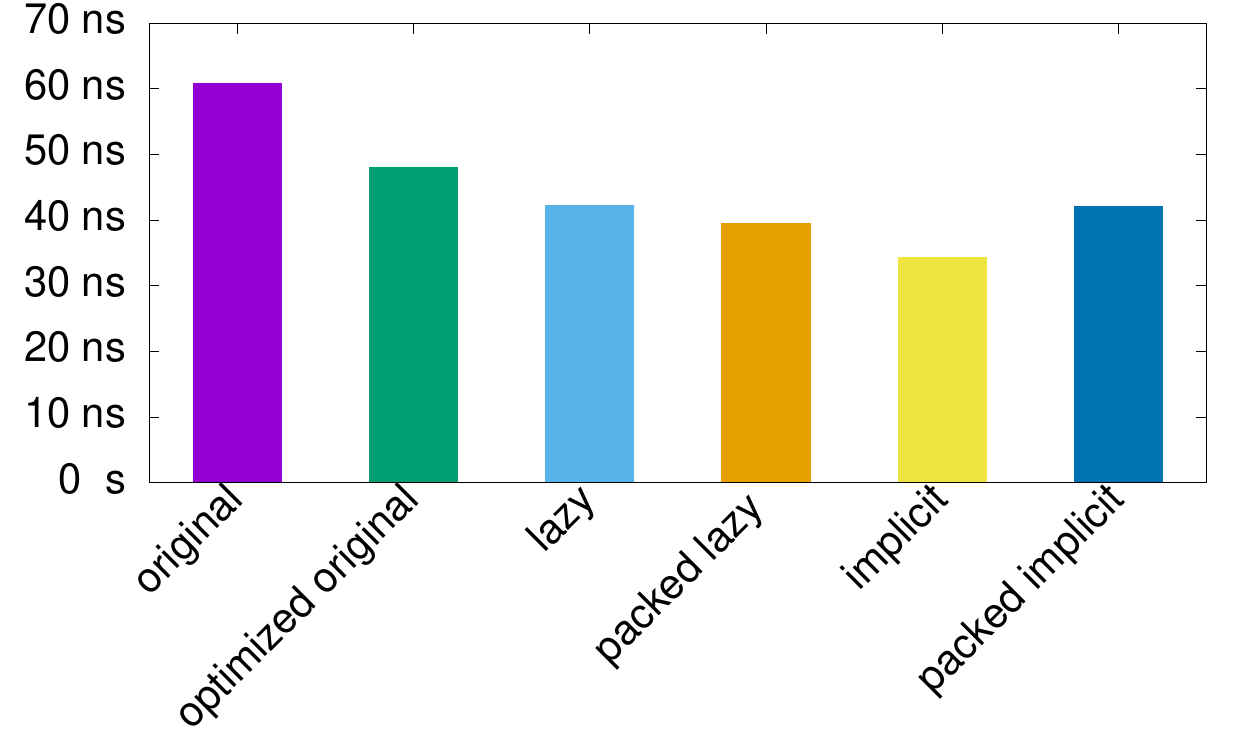}
		\caption{\textit{access}}
	\end{subfigure}
	\begin{subfigure}[b]{0.3\textwidth}
		\includegraphics[width=\textwidth]{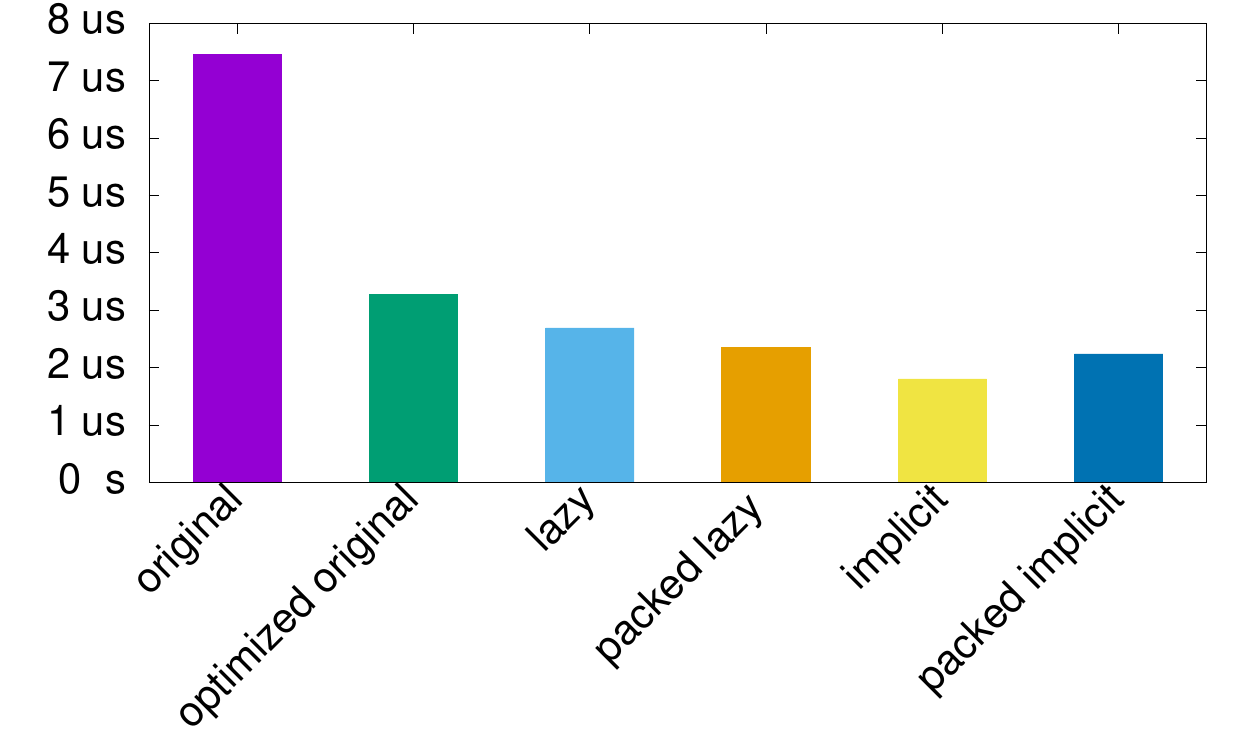}
		\caption{\textit{insert}}
	\end{subfigure}
        \caption{Figures (a) and (b) show the performance of the
            \textit{original} (\protect\purple), \textit{optimized original}
            (\protect\green), \textit{lazy} (\protect\blue) \textit{packed
            lazy} (\protect\orange),
            \textit{implicit} (\protect\yellow)
            and \textit{packed implicit} (\protect\dblue) layouts.}
\label{fig:test_representation}
\end{figure}
\subsection{Tiered Vector Variants}

In this test we compare the performance
of the implementations listed in Section~\ref{sec:implementation} to that 
or the original data structure as described in~\ref{thm:pointer}.

%\paragraph{Optimized Original}
\subparagraph*{Optimized Original}
By co-locating the child offset and child pointer, the two memory lookups are at
adjacent memory locations. Due to the cache lines in modern processors,
the second memory lookup will then often be answered directly by the fast
L1-cache.
As can be seen on Figure~\ref{fig:test_representation}, this small change in the memory layout results in a significant improvement in performance for both access and insertion. In the latter case, the running time is more than halved.

%\paragraph{Lazy and Packed Lazy}
\subparagraph*{Lazy and Packed Lazy}

Figure~\ref{fig:test_representation} shows
how the fewer memory probes required by the
\textit{lazy} implementation in comparison to the \text{original}
and \text{optimized original} results in better performance.
Packing the offset and pointer in the leaves results in even better performance
for both access and insertion even though it requires a few extra instructions
to do the actual packing and unpacking.

%\paragraph{Implicit}
\subparagraph*{Implicit}
From Figure~\ref{fig:test_representation}, we see the implicit
data structure is the fastest.
This is as expected because it requires fewer
memory accesses than the other structures except
for the packed lazy which instead has a slight
computational overhead due to the packing and unpacking.

As shown in Theorem~\ref{thm:implicit} the implicit data structure has a
bigger memory overhead than the lazy data structure.
Therefore the packed lazy representation might be beneficial in some
settings.

%\paragraph{Packed Implicit}
\subparagraph*{Packed Implicit}

Packing the offsets array could lead to 
better cache performance due to the smaller memory footprint and therefore
yield better overall performance.
As can be seen on Figure~\ref{fig:test_representation},
the smaller memory footprint
did not improve the performance in practice.
The simple reason for this,
is that the strategy we used for packing the offsets required
extra computation. This clearly dominated the possible gain from the
hypothesized better cache performance. We tried a few strategies to minimize
the extra computations needed at the expense of slightly worse memory usage,
but none of these led to better results than when not packing the offsets at
all.

\subsection{Width Experiments}

\begin{figure}
	\centering
	\begin{subfigure}[b]{0.3\textwidth}
		\includegraphics[width=\textwidth]{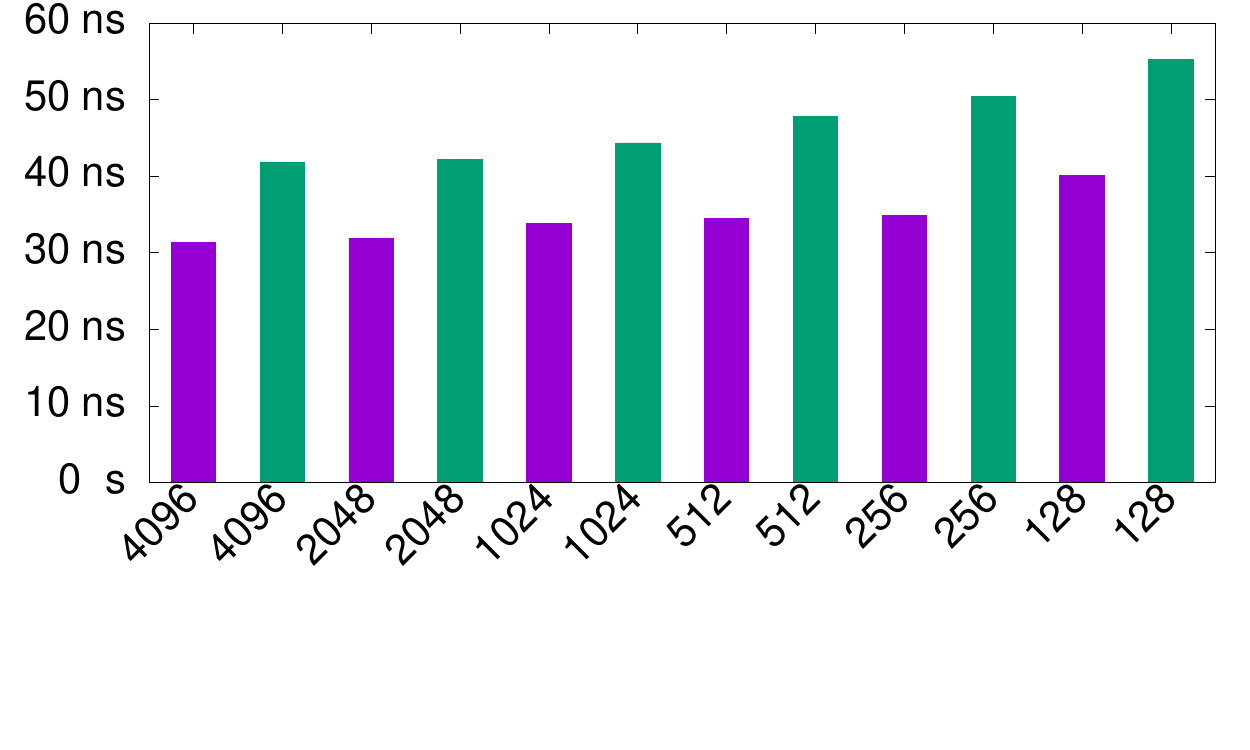}
		\caption{\textit{access}}
	\end{subfigure}
	\begin{subfigure}[b]{0.3\textwidth}
		\includegraphics[width=\textwidth]{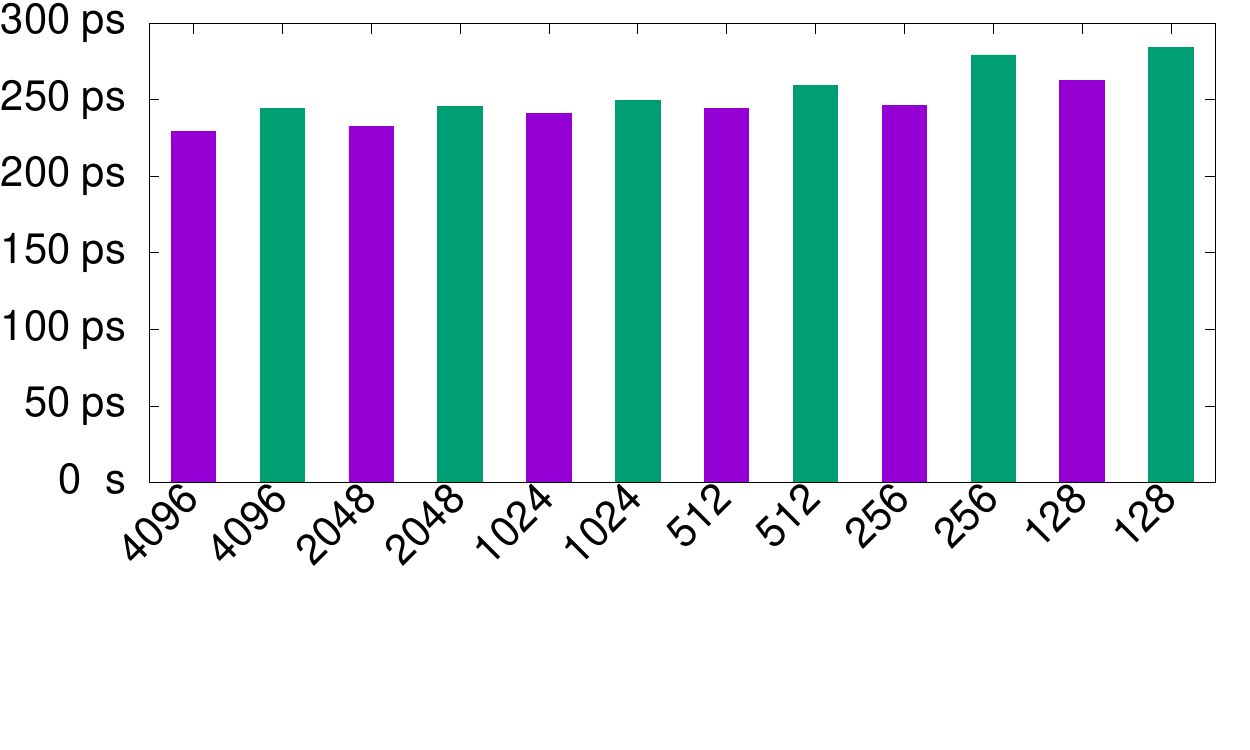}
		\caption{\textit{range access}}
	\end{subfigure}
	\begin{subfigure}[b]{0.3\textwidth}
		\includegraphics[width=\textwidth]{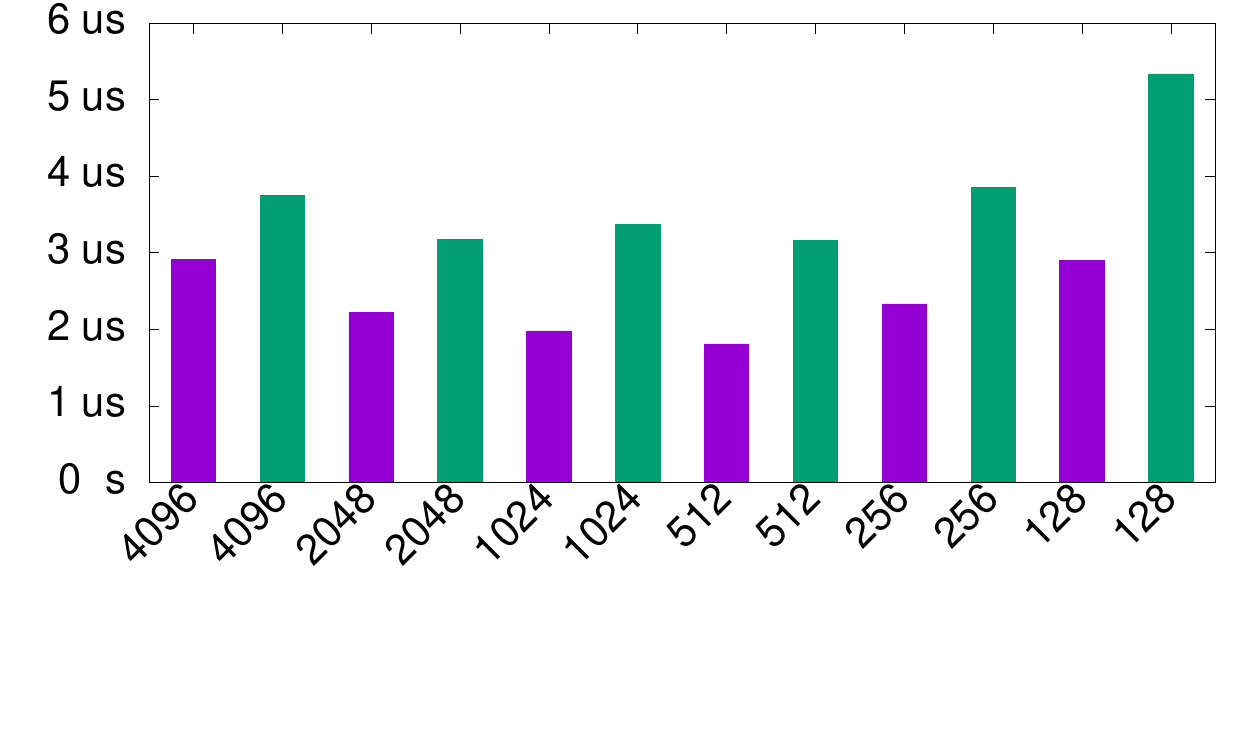}
		\caption{\textit{insert}}
	\end{subfigure}
	\caption{Figures (a), (b) and (c) show the performance of the \textit{implicit} (\protect\purple) and
		the \textit{optimized original} tiered vector (\protect\green) for different tree widths.}
\label{fig:test_width}
\end{figure}

This experiment was performed to determine the best capacity ratio between the leaf nodes and the internal nodes.
The six different width configurations we have tested are: 32-32-32-4096, 32-32-64-2048, 32-64-64-1024, 64-64-64-512, 64-64-128-256, and 64-128-128-128.
All configurations have a constant height 4 and a capacity of approximately 130 mio.

We expect the performance of access operations to remain unchanged, since the
amount of work required only depends on the height of the tree,
and not the widths. We expect range access to perform better when the leaf size
is increased, since more elements will be located in consecutive memory
locations. For $insertion$ there is not a clearly expected behavior as the time
used to physically move elements in a leaf will increase with leaf size, but
then less operations on the internal nodes of the tree has to be performed.

On Figure~\ref{fig:test_width} we see access times are actually decreasing
slightly when leaves get bigger. This was not expected, but is most likely
due to small changes in the memory layout that results in slightly better cache
performance. The same is the case for range access, but this was expected. For
insertion, we see there is a tipping point. For our particular instance, the
best performance is achieved when the leaves have size 512.

%Based on this, we have performed the remaining tests with the 64-64-64-512 configuration (unless otherwise specified).

\subsection{Height Experiments}

\begin{figure}
	\centering
	\begin{subfigure}[b]{0.3\textwidth}
		\includegraphics[width=\textwidth]{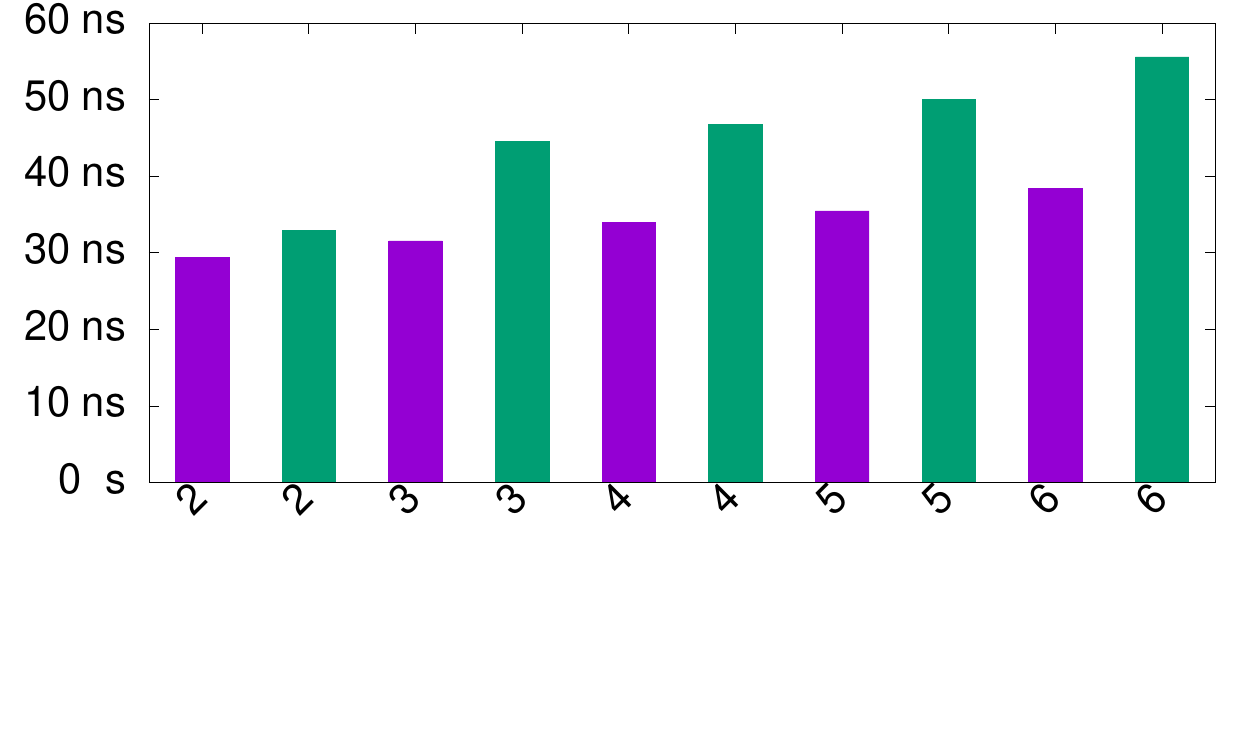}
		\caption{\textit{access(i)}}
	\end{subfigure}
	\begin{subfigure}[b]{0.3\textwidth}
		\includegraphics[width=\textwidth]{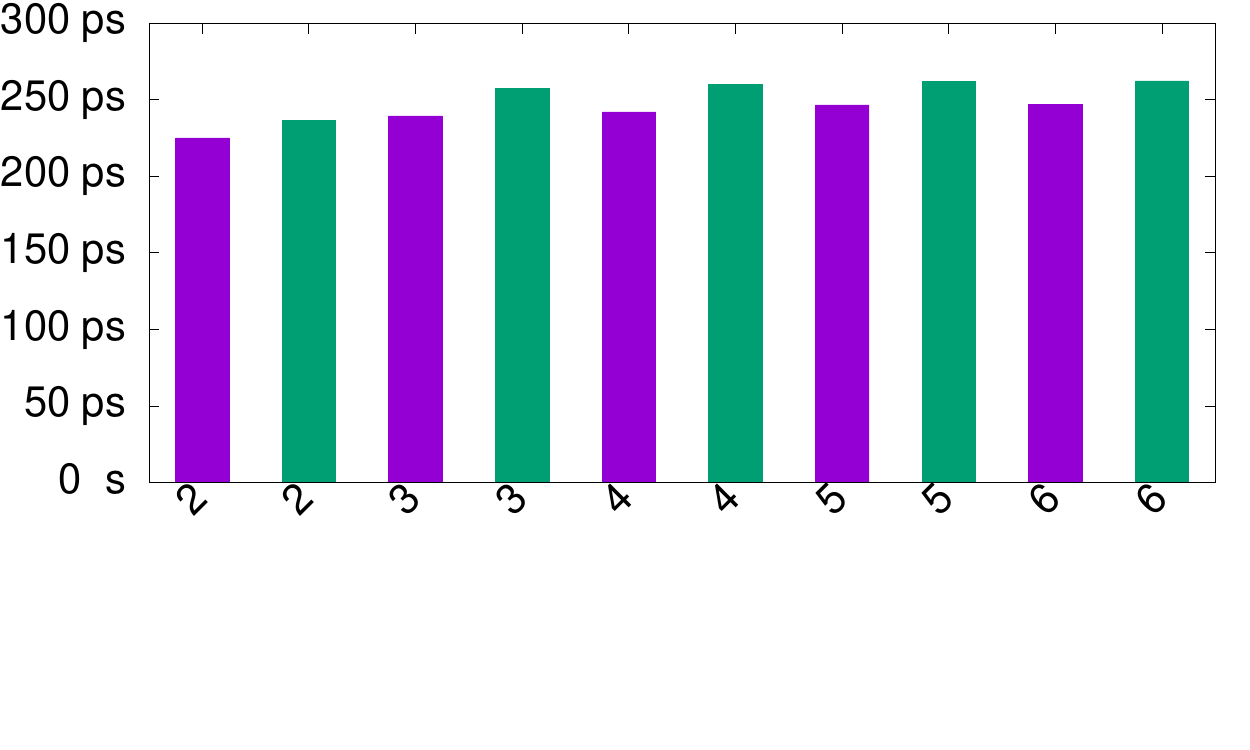}
		\caption{\textit{access(i, m)}}
	\end{subfigure}
	\begin{subfigure}[b]{0.3\textwidth}
		\includegraphics[width=\textwidth]{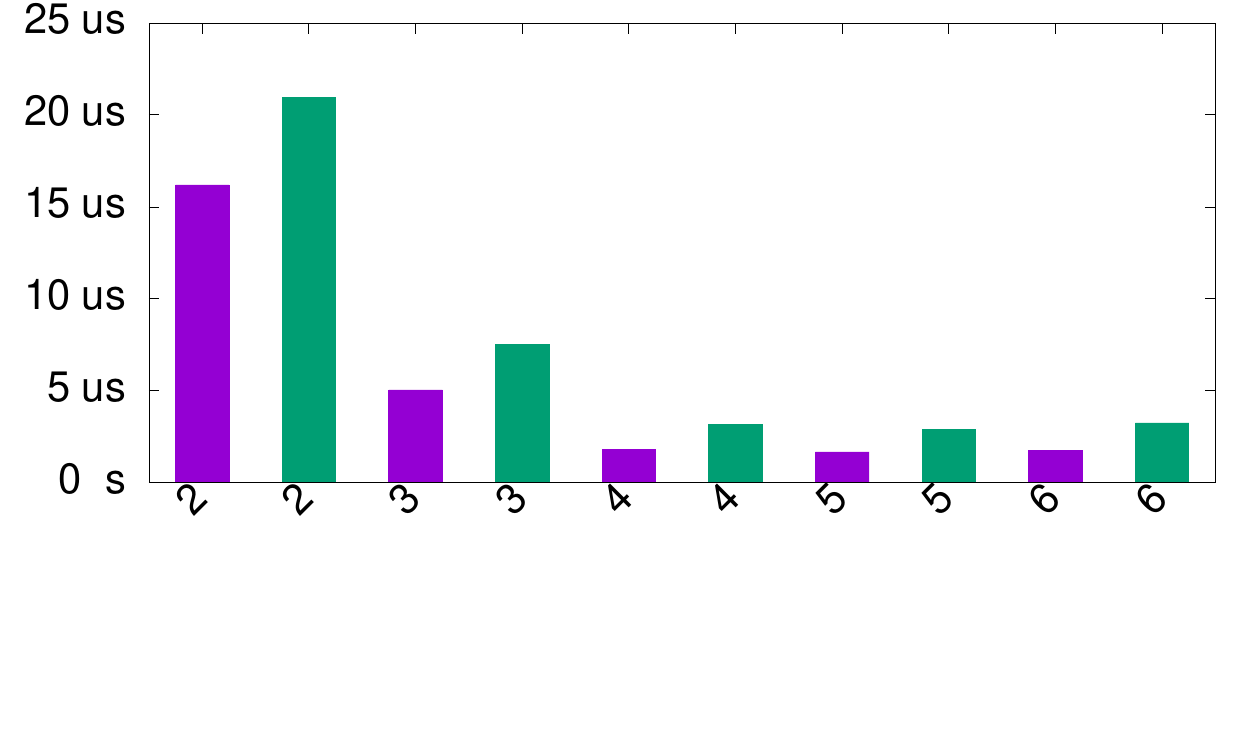}
		\caption{\textit{insert}}
	\end{subfigure}
	\caption{Figures (a),(b) and (c) show the performance of the \textit{implicit} (\protect\purple) and
		the \textit{optimized original} tiered vector (\protect\green) for different tree heights.}
\label{fig:test_height}
\end{figure}

In these tests we have studied how different heights affect the performance of
access and insertion operations. We have tested the configurations 8196-16384,
512-512-512, 64-64-64-512, 16-16-32-32-512, 8-8-16-16-16-512. All resulting in
the same capacity, but with heights in the range 2-6.

We expect the access operations to perform better for lower trees, since
the number of operations that must be performed is linear in the height. On the
other hand we expect insertion to perform significantly better with higher
trees, since its running time is $O(n^{1/l})$ where $l$ is the height plus one. 

On Figure~\ref{fig:test_height} we see the results follow our expectations. However, the access operations only perform slightly worse on higher trees.
This is most likely because all internal nodes fit within the L3-cache. Therefore the running time is dominated by the lookup of the element itself.
(It is highly unlikely that the element requested by an access 
to a random position would be among the small fraction of elements that
fit in the L3-cache).

Regarding insertion, we see significant improvements up until a height of 4. After that, increasing the height does not change the running time noticeably. This is most likely due to the hidden constant in $O(n^{1/l})$ increasing rapidly with the height.

\subsection{Configuration Experiments}

\begin{figure}
    \centering
    \begin{subfigure}[b]{0.3\textwidth}
        \includegraphics[width=\textwidth]{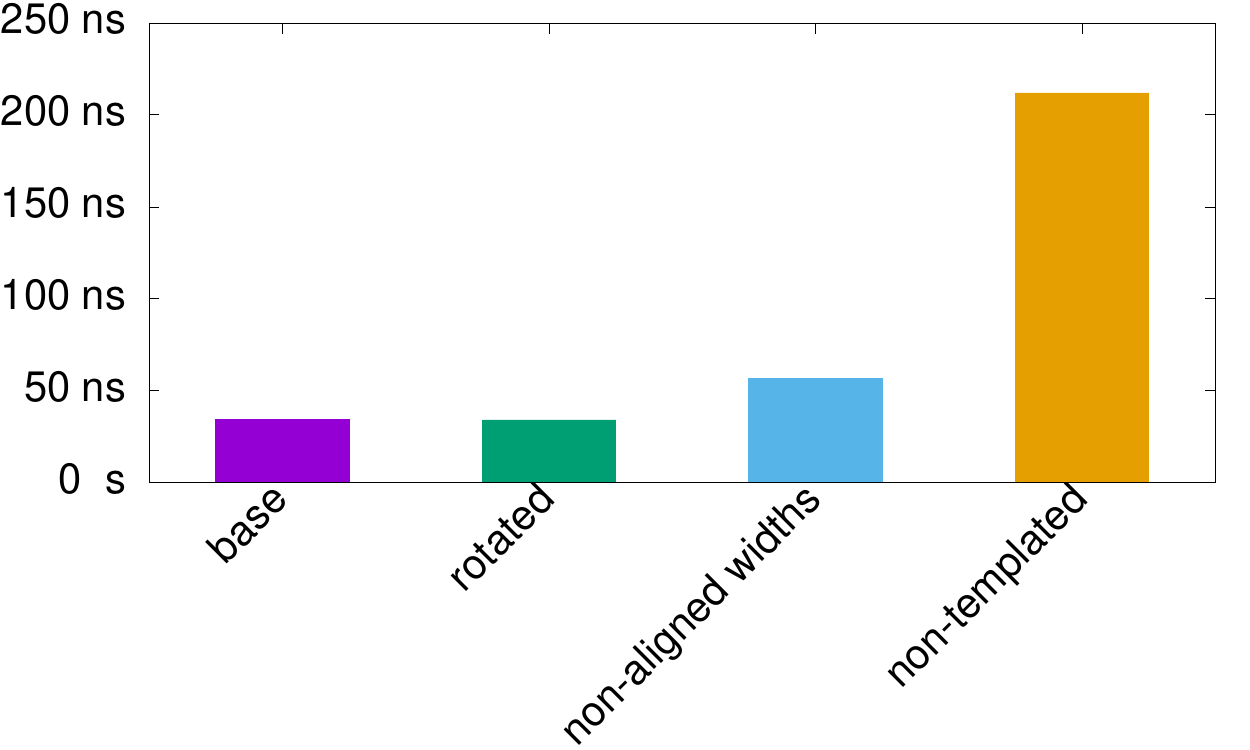}
        \caption{\textit{access}}
    \end{subfigure}
    \begin{subfigure}[b]{0.3\textwidth}
        \includegraphics[width=\textwidth]{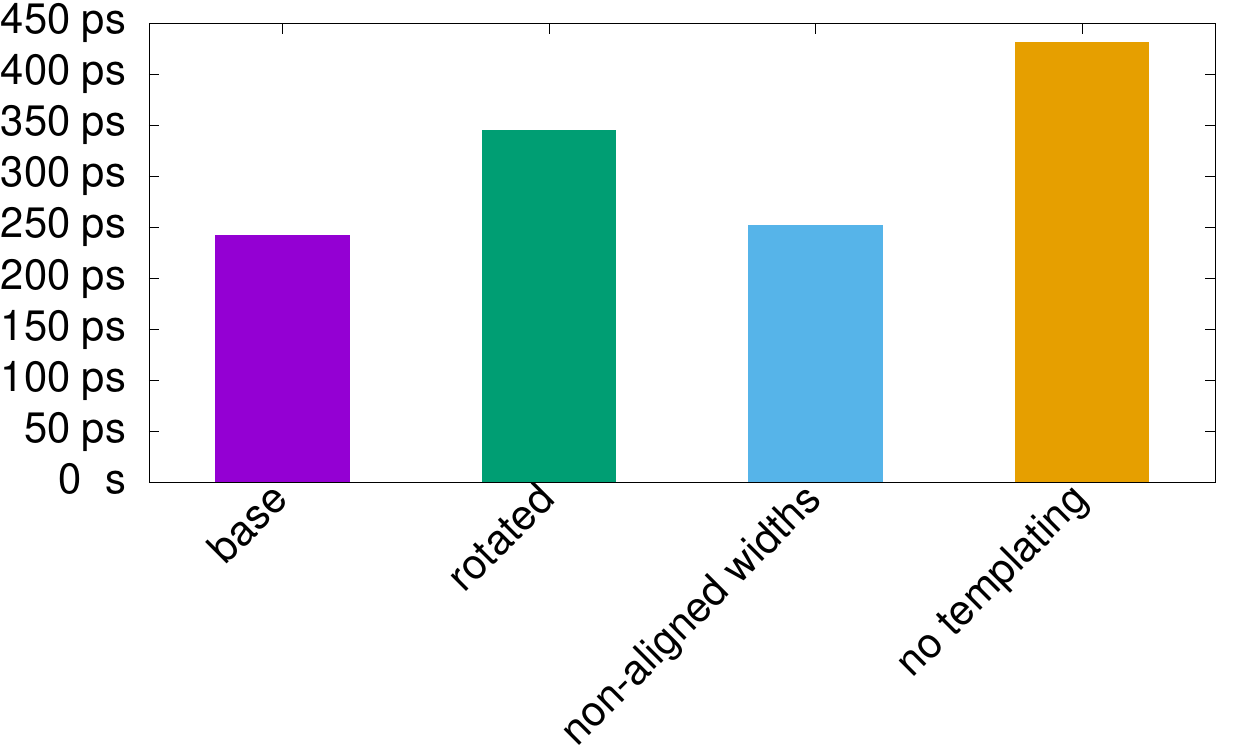}
        \caption{\textit{range access}}
    \end{subfigure}
    \begin{subfigure}[b]{0.3\textwidth}
        \includegraphics[width=\textwidth]{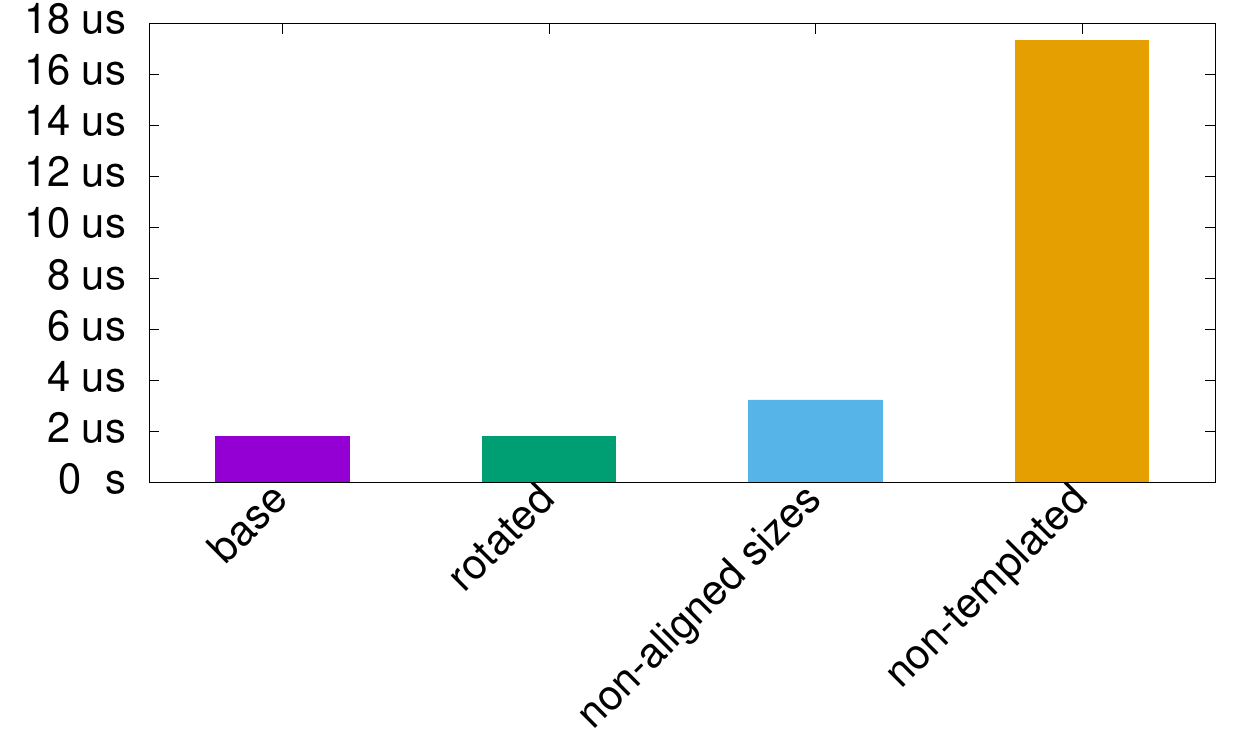}
        \caption{\textit{insert(i,x)}}
    \end{subfigure}
    \caption{Figures (a) and (b) show the performance of the
    \textit{base} (\protect\purple),
    \textit{rotated} (\protect\green), 
    \textit{non-aligned sizes} (\protect\blue),
    \textit{non-templated} (\protect\orange)
    layouts.}
\label{fig:test_minor}
\end{figure}

In these experiments, we test a few hypotheses about how different changes
impact the running time. The results are shown on
Figure~\ref{fig:test_minor}, the leftmost result (base) is
the implicit 64-64-64-512 configuration of the tiered vector 
to which we compare our hypotheses.
%our final and best

\textit{Rotated}: 
As already mentioned, the insertions performed as a
preliminary step to the tests are not done at random positions.
This means that all offsets are zero when our real operations
start. The purpose of this test is the ensure that
there are no significant performance gains in starting
from such a configuration which could otherwise
lead to misleading results.
To this end, we have randomized all
offsets (in a way such that the data structure is still valid, but the
order of elements change) after doing the preliminary insertions
but before timing the operations. As can be seen on
Figure~\ref{fig:test_minor}, the difference between this and the normal
procedure is insignificant, thus we find our approach gives a fair picture.

\textit{Non-Aligned Sizes}: In all our previous tests, we have ensured all
nodes had an out-degree that was a power of 2. This was chosen in order to let the
compiler simplify some calculations, i.e.\ replacing multiplication/division
instructions by shift/and instructions. As Figure~\ref{fig:test_minor} shows,
using sizes that are not powers of 2 results in significantly worse performance.
Besides showing that powers of 2 should always be used, this also indicates that not only
the number of memory accesses during an operation is critical for our
performance, but also the amount of computation we make.

\textit{Non-Templated}
The non-templated results 
in Figure~\ref{fig:test_representation} the
show that the change to templated recursion
has had a major impact on the running time. It should be noted that some
improvements have not been implemented in the non-templated version,
but it gives a good indication that this has been quite useful.

\section{Conclusion}

This paper presents the first implementation of a generic tiered vector
supporting any constant number of tiers.
We have shown a number of modified versions of the tiered vector, and employed several optimizations to the implementation. These implementations have been compared
to vector and multiset from the C++ standard library.
The benchmarks show that our implementation
stays on par with vector for access and on update operations
while providing a considerable speedup of more than $40\times$ compared to multiset. 
At the same time the asymptotic difference between the logarithmic complexity
of multiset and the polynomial complexity of tiered vector
for insertion and deletion operations only has little effect in practice.
For these operations, our fastest version of the tiered vector suffers less than 10\% slowdown.
Arguably, our tiered array provides a better trade-off
than the balanced binary tree data structures used in the standard library for most applications
that involves big instances of the dynamic array problem.

\bibliography{references}

\end{document}